\documentclass[aps,pra,reprint]{revtex4-1}
\usepackage{blindtext}
\usepackage{graphicx}
\usepackage{epstopdf}
\usepackage{amsmath}
\usepackage{amsmath,mathtools}

\usepackage{lipsum}

\DeclareMathOperator\erf{erf}

\begin{document}

\title{Effects of quantum fluctuations on the dynamics of dipolar Bose polarons}

\author{Nadia Guebli and Abdel\^{a}ali Boudjem\^{a}a}

\affiliation{Department of Physics, Faculty of Exact Sciences and Informatics, Hassiba Benbouali University of Chlef P.O. Box 78, 02000, Ouled Fares, Chlef, Algeria.}



\begin{abstract}

We study the dynamics of dipolar Bose polarons in the presence of the normal and anomalous fluctuations using the time-dependent Hartree-Fock-Bogoliubov theory.
The density profiles of the condensate, the anomalous component and the impurity are deeply analyzed.
The time evolution of the width and the center-of-mass oscillation of such quantities is also highlighted.
We calculate corrections due to quantum fluctuations and impurity to the chemical potential and the radius of the condensate and of the anomalous component 
in the weak coupling regime using the Thomas-Fermi approximation.
Effects of the dipole-dipole interaction, impurity-host interaction and the anomalous fluctuation on the width and on the breathing frequencies of the impurity are 
discussed by variational and numerical means. 

\end{abstract}

\maketitle

\section{Introduction}

Ultracold atomic impurities immersed in a Bose-Einstein condensate (BEC), known as Bose polaron problem
has attracted growing attention in recent years in many areas of physics, such as condensed matter physics 
quantum atomic gases and quantum information.

Experimentally, the impurity and its atomic bath have been realized in the strongly interacting regime in one-dimensional  (1D)  \cite{Cat1} 
and 3D \cite{Guang, Jorg} geometries.
Theoretically, BEC-impurity mixtures have been intensively investigated using many approaches \cite{Dev, Grud}.
Nevertheless, most theories describing Bose polarons, were based on the effective Fr\"ohlich model \cite {Brud1, Temp, Cast, Cast1,  Shash, Grus, Vlt, Shcha}
within different many-body techniques.  The generalized Hartree-Fock-Bogoliubov (HFB) description of the Fr\"ohlich method  has been derived in Ref.\cite {Kain1}.
Other good theoretical tools, including a mean-field theory to examine the self-localization effects \cite {Astr, Tim1, Tim2, Blum, Brud, Jian, Vol},
a third-order perturbative treatment of the problem \cite {Ras}, a variational approach \cite {Jesp}, a self-consistent $T$-matrix calculation \cite {Rath}, 
and numerical simulations based on the quantum Monte Carlo method \cite {Pen, Pars, Grus, Grus1}.
The universality of such Bose polarons has been adressed in Ref.\cite{Shu} using a variational approach. 
Furthermore, temperature effects on BEC-impurity mixture have been discussed utilizing the time-dependent HFB theory (TDHFB) \cite {Boudj,Boudj1, Boudj2, Boudj3}, 
diagrammatic approach \cite {Sun, Levin, NE}, and spectroscopy of Rydberg excitations \cite{Schmd}.
At low temperatures, the properties of a polaron in dilute 2D Bose gas have been discussed in \cite{Pastu} in a weakly-coupled regime using a perturbative theory. 

However, the experimental achievement of BEC with dipole-dipole interaction (DDI) \cite{Baranov, Pfau} paves the way towards a new exciting research area, 
namely dipolar polarons i.e. impurity-BEC mixtures with DDI.
The Bogoliubov phonon modes and the ground-state properties of such systems have been discussed in Refs.\cite{Kain, Pena} within the framework of the Fr\"ohlich model.  
Effects of DDI, impurity-bosons interaction and external trapping potentials
on the breathing modes of impurity have been analyzed using a variational scheme \cite{FQ}.
Furthermore, the depletion due to the impurities of dipolar Bose polarons has been examined using an effective action in the mean-field approximation
and found to be independent of the condensed density \cite{Must}. 

While most of the previous works give good results, many questions remain unanswered, for instance, effects of quantum normal and anomalous fluctuations
on the equilibrium and on the dynamics of dipolar polarons.
It has been shown that the presence of the anomalous density (pairing correlations) in quantum ultracold gases leads to novel physical effects, 
in particular at intermediate temperatures \cite{Griffin, Burnet, Yuk, Boudjbook}.

In this paper we investigate the behavior of a harmonically trapped dipolar Bose polarons in the presence of the normal and anomalous correlations 
and demonstrate how this latter manifests itself in determining the static and dynamical properties of the impurity and its host.
To this end, we use our TDHFB theory  which is a powerful tool for  exploring many-particle systems and Bose polarons \cite{Boudj, Boudj1, Boudj2, Boudj3}.
The theory is valid for arbitrary temperatures and densities as well as it properly takes into account the dynamics of the pairing correlations 
\cite{Boudj,Boudj1,Boudj2,Boudj3, Boudjbook, Boudj4,Boudj5, Boudj6, Boudj7, Boudj8, Boudj9, Boudj10, Boudj11, Boudj12}.
The TDHFB theory governs selfconsistently the dynamics and the static of the condensate, impurity, thermal cloud and the anomalous density.
It is based on the time-dependent Balian-V\'en\'eroni variational principle which optimizes 
the evolution of the state according to the relevant observable in a given variational space \cite{BV}. 
Recently, the TDHFB theory has been revisited with renormalized coupling constant and more accurate description of dynamical and thermodynamic properties of BEC
has been achieved \cite{Boudjbook, Boudj7, Boudj8, Boudj9, Boudj10, Boudj11}.

The aim of the present work is to investigate the breathing modes of the condensate, the impurity and the anomalous component for 
both repulsive and attractive impurity-boson interactions. 
We show that the presence of the pairing fluctuations, the dipolar and interspecies interactions, and external trapping potentials  
may crucially change the density profiles and the breathing oscillations of such quantities. 
Generalized analytical expressions for chemical potential and Thomas-Fermi (TF) radii of the condensate are derived in terms of the condensed and anomalous fractions.
By means of the variational approach, we calculate the impurity's effective potentials, the widths and the breathing modes in the weak coupling regime.
We find that the DDI, impurity-bath interaction, and pairing correlations may modify 
the oscillation amplitude of the impurity's size and the frequencies of the breathing modes.

The rest of the paper is organized as follows. In Sec.\ref{Mod}, we illustrate the TDHFB model and establish its validity for the problem at hand.  
We analyze the density profiles predicted by a direct numerical simulation of our TDHFB  equations for the condensate, the impurity and the anomalous component.
The dynamics of the condensate and the anomalous density in the presence of the impurity is also discussed.
In the weak coupling regime, generalized analytic expressions for the chemical potential and condensate radii
are obtained within the framework of the TF approximation.
The intent of Sec.\ref{SLS} is to analyze the impact of the impurity-BEC  interaction strength, the DDI and the pairing correlations on the
time evolution of the width and the breathing modes of a harmonically trapped impurity interacting with a trapped BEC.
To this end, we utilize a variational method based on our TDHFB equations.
Our results are compared by the findings computed by the HFB-Popov approximation.
Finally, we conclude this paper in Sec.\ref{Conc}.

\section{TDHFB theory} \label{Mod}

We consider a single impurity atoms of mass $m_I$ in a cylindrically symmetric harmonic trap $U_I =(m_I \omega_{I \rho}^2/2) (\rho^2 + \lambda_I^2 z^2)$, 
immersed in identical dipolar bosons of mass $m_B$ trapped by an external potential $U_B =(m_B \omega_{B \rho }^2/2) (\rho^2+ \lambda_B^2 z^2)$,  
where $\lambda_B=\omega_{B z}/\omega_{B \rho}$ and $\lambda_I=\omega_{I z}/\omega_{I \rho}$ with $\omega_\rho$ and $\omega_z$  
being respectively, the radial and axial frequencies of  bosons trapping potential.
The boson-boson and impurity-boson interactions are approximated by contact potentials: $g_B \delta ({\bf r}-{\bf r'})$ and $g_{IB} \delta ({\bf r}-{\bf r'})$, 
where $g_B=(4\pi \hbar^2/m_B) a_B$ and $g_{IB}=2\pi \hbar^2 (m_B^{-1}+m_I^{-1}) a_{IB}$ with 
$a_B$ and $a_{IB}$ being the boson-boson  and  impurity-boson scattering lengths, respectively. 
The DDI between two bosons is $V_d({\bf r}-{\bf r'}) = C_{dd} (1-3\cos^2\theta) / (4\pi |{\bf r}-{\bf r'}|^3)$,
where the coupling constant $C_{dd} $ is ${\cal M}_0 {\cal M}^2$ for particles having a permanent magnetic dipole moment ${\cal M}$ (${\cal M}_0$ is the magnetic permeability
in vacuum) and $d^2/\epsilon_0$ for particles having a permanent electric dipole $d$ ($\epsilon_0 $ is the permittivity of vacuum),
and $\theta$ is the angle between the relative position of the particles ${\bf r}$ and the direction of the dipole.
In our case the dipole moment of the atoms are assumed to be directed along the $z$-direction.

The TDHFB equations governing the dynamics of the condensate, thermal cloud, the anomalous density and the impurity read  \cite{Boudj,Boudj1,Boudj2,Boudj3, Boudj11,Boudj12}
\begin{subequations}\label{E:eq4}
\begin{align}
i\hbar \dot{\Phi}_B & = \left [h_B^{sp} +g_B \left(n_c+2\tilde n +\gamma \, n_I \right)+ \Psi_{dd} \right]\Phi_B \label{E:eq4a} \\
&+ g_B \tilde m \Phi_B^*,  \nonumber\\ 
i\hbar \dot{\tilde{m}}& = 4\left\{h_B^{sp}+g_B \left[ 2 n_B  + \frac{1}{4} \left (2\tilde n+1 \right)+ \gamma n_I \right] \right\} \tilde{m} \nonumber \\
& +g_B \left (2\tilde n+1 \right) \Phi_B^2, \label{E:eq4b} 
\\
i\hbar \dot{\Phi}_I & = \left (h_I^{sp} +g_B\gamma \, n_B \right) \Phi_I,  \label{E:eq4c} 
\end{align}
\end{subequations}
where $\Phi_B$ and $\Phi_I$ stand for the condensate and the impurity wavefunctions, respectively. 
The quantities  $n_c=|\Phi_B|^2$ and $n_I=|\Phi_I|^2$ represent, respectively the densities of the condensate and  the impurity.
The noncondensed $\tilde n $ and the anomalous density $\tilde m$ are identified, respectively with $\langle \hat {\bar \psi}_B^{\dagger} \hat {\bar \psi}_B\rangle$ 
and $\langle \hat {\bar \psi}_B \hat {\bar \psi}_B\rangle$ with $\hat{\bar \psi}_j({\bf r})=\hat\psi_j({\bf r})- \Phi_j({\bf r})$ being the noncondensed part of the field operators 
$\hat\psi_j(\bf r)$ ($j$ stands for the boson and impurity).
The total density in BEC is defined by $n_B = n_c+ \tilde n$.
The single particle Hamiltonian for the condensate and the impurity are defined, respectively as
$h_B^{sp}= -(\hbar^2/ 2m_B) \Delta + U_B$ and $ h_I^{sp}= -(\hbar^2/ 2m_I) \Delta + U_I$. 
The DDI is described by the term $\Psi_{dd}  ({\bf r}) =\int d {\bf r'}  V_d ({\bf r}-{\bf r'}) |\Phi_B ({\bf r'})|^2 $.
The ratio $\gamma=g_{IB}/g_B$ controls the relative strengths of interactions.
It is worth noticing  that Eqs.(\ref{E:eq4}) are derived under the assumption that the long-range exchange terms   
$\tilde n ({\bf r},{\bf r'})=\tilde m ({\bf r},{\bf r'})=0$ for ${\bf r} \neq {\bf r'}$ \cite{Bon,Bon1},
since they do not influence the stability of the system \cite {Bon, He, Biss, Zhan, Bail, Tick, Boudj5, Boudj6}.     
In the absence of the DDI, Eqs.(\ref{E:eq4}) reduce to those of Bose polarons with pure contact interactions \cite{Boudj,Boudj1,Boudj2,Boudj3}.
For $\tilde{m}=0 $, one recovers the well-known HFB-Popov equation. 
Setting $g_{IB}=0 $ and  $\tilde{m}=0$, Eqs.(\ref{E:eq4}) reproduce  the Gross-Pitaevskii (GP) equation 
describing a degenerate dipolar Bose gas at zero temperature and the usual Schr\"{o}dinger equation describing a non-interacting impurity system.  
Stationary states corresponding to Eqs.(\ref{E:eq4}) can be obtained using:
$\Phi_B ({\bf r},t)=\Phi_B  ({\bf r}) \exp (-i\mu_B t/\hbar)$, $\tilde{m}  ({\bf r},t)=\tilde{m}  ({\bf r})\exp (-i\mu_B t/\hbar)$, 
and $\Phi_I ({\bf r},t)=\Phi_I ({\bf r}) \exp (-i\mu_I t/\hbar)$, where $\mu_B$ and $\mu_I$ are, 
respectively the chemical potentials of the bosonic and impurity components.

In our TDHFB formalism, equations (\ref{E:eq4}) are complemented by an equation connecting the normal and anomalous densities (see e.g.\cite{Boudjbook}) 
\begin{equation}  \label{Inv1}
\tilde n (\tilde n+1)=|\tilde m|^2.
\end{equation}
Equation (\ref{Inv1}) which is valid at zero temperature renders the system (\ref{E:eq4}) close.
It shows that the anomalous density is larger than the noncondensed density even at very low temperature signaling the 
importance of $\tilde m$ in ultracold Bose gases.

It is well known that the inclusion of the anomalous density which is ultraviolet divergent quantity leads to the appearence of an unphysical gap 
in the spectrum. To solve such a problem, one should use the renormalized coupling constant  $\bar g_B= g_B (1+\tilde m/\Phi_B^2)$
\cite{Boudj, Boudj1, Boudj2, Boudj3, Boudj11, Burnet, Boudjbook}.
The spatially dependent effective interaction $\bar g_B $ plays a crucial role in stabilizing the system (rendering the spectrum gapless)
and has to be self-consistently determined together with the condensate wavefunction and the anomalous density.
The whole TDHFB equations are thus solved iteratively.


\subsection{Density profiles} \label{DP}

To earn more specific insights into the dynamics of dipolar polarons, 
we solve our TDHFB equations (\ref{E:eq4}), numerically following the procedure described in our recent works \cite{Boudj4, Boudj7}. 
Lengths, time and energies are expressed in terms of the transverse harmonic oscillator length $l_B=\sqrt{\hbar/m \omega_{\rho}}$,
the transverse trapping frequency is $\omega_{B \rho}^{-1}$, and the trap energy $\hbar \omega_{\rho}$, respectively. 
We introduce also  the dimensionless quantity $\varepsilon_{dd}= C_{dd}/3g_B$ which represents the relative strength between the DDI and the contact interaction.
We set the ratio mass $\alpha=m_B/m_I$ and $\Omega_{\rho}=\omega_{B \rho}/\omega_{I \rho}$. 

\begin{figure} 
\centerline{
\includegraphics[scale=0.45]{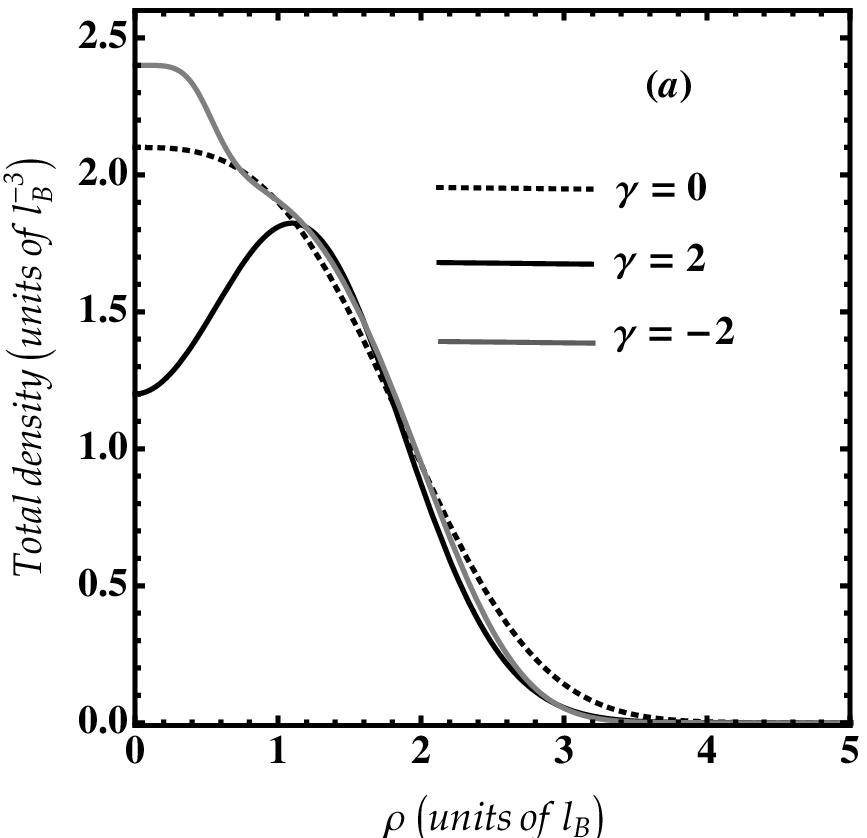}
\includegraphics[scale=0.45]{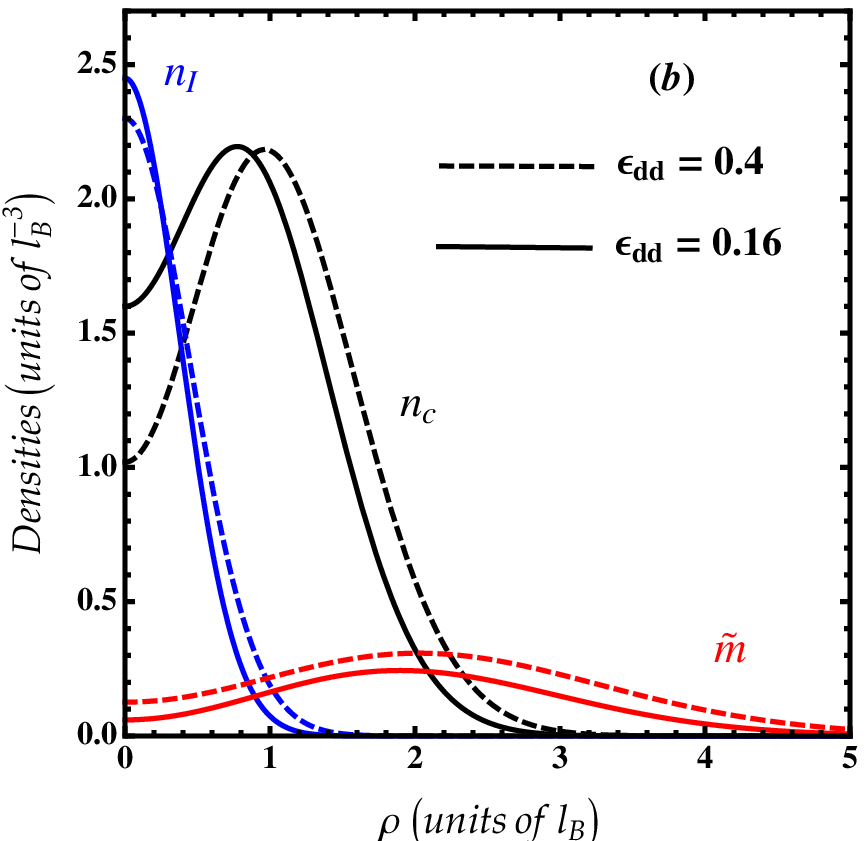}}
 \caption{ (Color online)  (a) Deformation of the total radial density for both repulsive and attractive impurity-host interactions for $\epsilon_{dd}=0.4$ (corresponding ${}^{166}$ Er atoms).
(b) Density profiles for a dipolar BEC-impurity mixture for $\gamma=2$.
Parameters are: $N=10^4$, $a_B= 100 \,a_0$ with $a_0$ being the Bohr radius, $\omega_{B \rho}=2\pi \times 500$ Hz, $\lambda_B=\lambda_I=5$,
$\alpha=2$, and $\Omega_{\rho}=0.2$.}
\label{DF} 
\end{figure}

Figure \ref{DF}.a depicts the role of a single impurity atom in the behavior of the total density of a dipolar BEC  for both repulsive and attractive 
impurity-host interactions.
We see that for $\gamma >0$, the impurity is localized inside the BEC and becomes 
self-trapped  while for $\gamma < 0$, the density of the BEC is considerably increased (peaked) as in the nondipolar case \cite{Boudj1, Boudj2}.
Note that the impurity-host interaction is sorely tunable utilizing a Feshbach resonance \cite{Cat1,Guang, Jorg, Dev, Grud}.

At zero temperature the condensed density is comparable to the total density and both quantities are distorted by the impurity as is shown in Fig.\ref{DF}.b.
By further increasing the DDI, the depth of the dip formed by the impurity near the center of the trap is increasing 
owing to the long-range repulsive force exerted by the dipoles on the one hand and the inherent repulsive impurity-bath forces on the other.
We see also that the radius of the condensate is broadened with DDI.
In opposite, the anomalous density increases with DDI which is in fact natural since such a quantity survives and grows with interactions.
The presence of the impurity leads also to modify the shape of  $\tilde m$ as is seen in the same figure.

\begin{figure}
\includegraphics[scale=0.46]{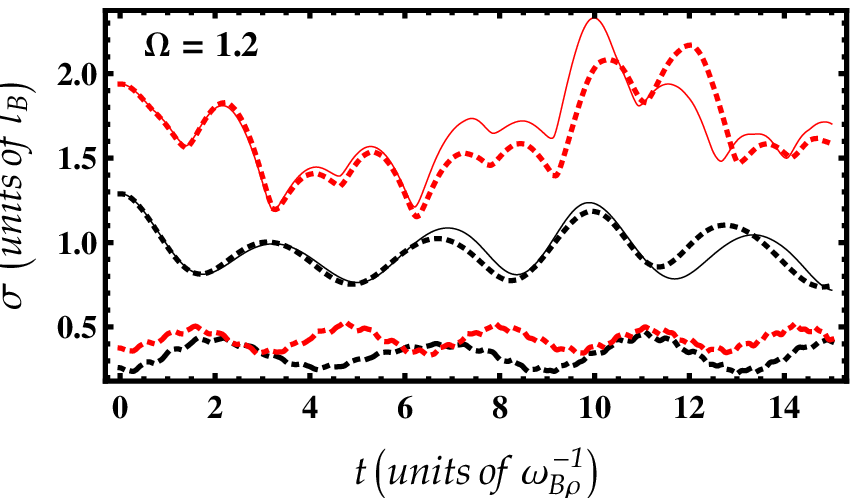}
\includegraphics[scale=0.46]{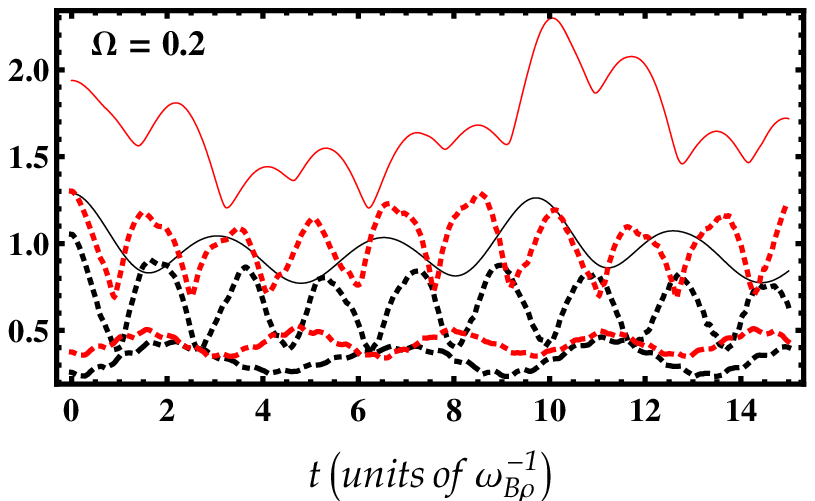}
\includegraphics[scale=0.46]{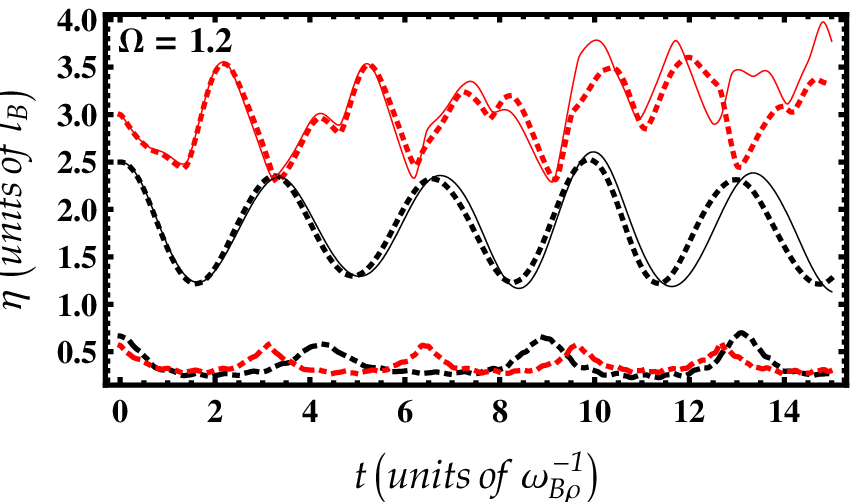}
\includegraphics[scale=0.46]{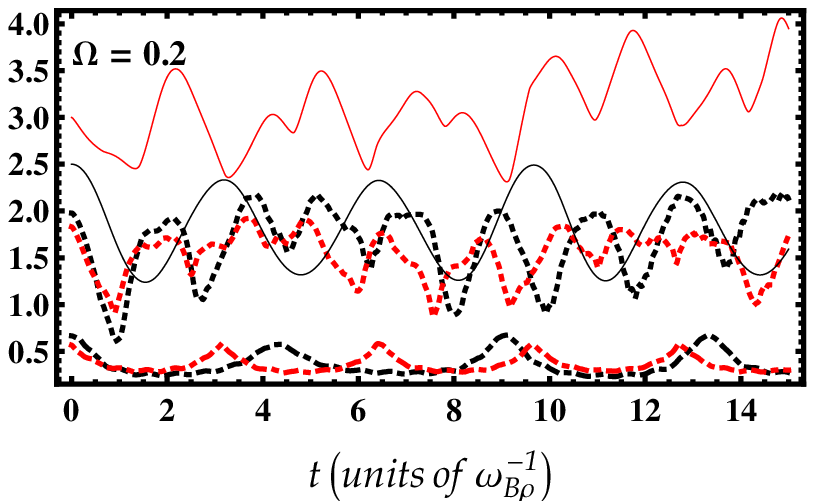}
\caption{(Color online) Dynamic evolution of the radial width (top panel) and radial center-of-mass (bottom panel) of the
condensate (black lines) and the anomalous component (red lines) for $\epsilon_{dd}=0.4$.
Dotted lines: $\gamma=0$.  Solid lines: $\gamma = 2$. Dotdashed lines: $\gamma = -2$. Parameters are the same as in Fig.\ref{DF}.}
 \label{CW}
\end{figure}

Now we turn to analyze the time evolution of the width and the center-of-mass of both the condensate and  the anomalous component
following the procedure reported in the experiment by Catani et. {\it al} \cite{Cat1}. 
The idea is that the impurity is initially trapped in the potential of axial and radial frequencies and the impurity-BEC mixture is cooled to the ground state \cite{Cat1, FQ,THJ}. 
At time $t=0$, the axial trap frequency is reduced and the subsequent evolution of the width for $t \geq 0$ is observed \cite{Cat1, FQ, THJ} 
(the width moves with lower trapping frequency).
The stationary solutions of our equations (\ref{E:eq4}) are used as initial conditions for handling the dynamics of the system.
Numerically, the width and the center-of-mass can be  extracted employing : $\sigma_i=\sqrt{ \int r ^2 |\Psi_i ({\bf r},t)|^2 d {\bf r}}$, and 
$\eta_i=\int  r |\Psi_i ({\bf r},t)|^2 d {\bf r} $, respectively, where $\Psi_i ({\bf r},t)$ stands for  $\Phi_B ({\bf r},t)$, $\tilde m ({\bf r},t)$, and $\Phi_I ({\bf r},t)$
and $i=z, \rho$. 
Plots of such quantities for different values of  $\gamma$, and $\Omega$ are shown in Fig.\ref{CW}. 

Periodic (breathing) oscillations are observed in the condensate which may continue for ever with different frequencies in the radial and axial directions. 
In the case of repulsive impurity-host interactions ($\gamma=2$) and for $\Omega_{\rho}=1.2$, we see that a single impurity may slightly enhance the width 
and the center-of-mass of  both the condensate and the anomalous component only at larger times.
Whereas, for attractive impurity-BEC interactions ($\gamma=-2$),  the presence of the impurity can produce a significant decrease in 
oscillation amplitudes of the width and the center-of mass of $n_c$ and $\tilde m$.

Remarkably, for small trapping frequency ($\Omega_{\rho}=0.2$),  amplitudes of oscillations of the width and the center-of-mass for a dipolar BEC with and without
impurity are found to differ substantially regardless the type of interspecies interactions. 
This discrepancy can be attributed to the interplay between trap anisotropy and dipolar interactions.
By increasing the DDI, the amplitude modulation of oscillations of $n_c$ and $\tilde m$ becomes stronger.
One can expect that the dipole modes of the impurity inside the bath can be shifted due to the DDI and interspecies interaction.
In order to experimentally observe the impurity one needs to tune $g_{IB}$ before the particle release, otherwise it remains localized inside the condensate.

\subsection{Thomas-Fermi approximation} \label{TFA}

In the limit of strong interactions or large number of particles, one can apply the so-called TF approximation 
which requires that the kinetic terms in Eqs.(\ref{E:eq4a}) and (\ref{E:eq4b}) should be neglected.
The TF equations then read 
\begin{subequations}\label{TF:eq}
\begin{align}
n_c& = n_{c0} -\gamma G n_I,   \label{TF:eq1}
\\
\tilde m&=\tilde m_0-\gamma (\beta-2) n_I,   \label{TF:eq2}
\end{align}
\end{subequations}
where the densities $n_{c0} = (\mu_B -U_B- a_2 \Psi_{dd})/g_B a_1$ and $\tilde m_0=(\mu_B -U_B- a_4 \Psi_{dd})/g_B a_3$.
The parameters $a_1$, $a_2$, $a_3$, and $a_4$ are defined respectively as: $a_1=(\beta+G\beta-2)/G$, $a_2=(G+1)/G$, $a_3=(\beta+G\beta-2)/(\beta-2)$, and $a_4=2/(\beta-2)$
with $\beta=\bar g_B /g_B$, and $G=\beta/4(\beta-1)$ being effective interaction strengths. 
One should stress that for large $\beta$, the system becomes strongly correlated. 
Therefore, the diluteness of the gas requires the condition: $\tilde m/\Phi^2 \ll 1$ or $\beta \rightarrow 1$ \cite{Boudj,Boudj1,Boudjbook}.
The dipolar potential inside the inverted parabola can be written as \cite{Dell,Dell1}
\begin{align} \label{DDIp}
\Psi_{dd}&= n_0 g_B\bigg[ \left(\frac{1}{\kappa^2}+\frac{3 f(\kappa)}{2(\kappa^2-1)}\right)\frac{\rho^2}{ R_\rho^2}  \\ 
&-\left(2+\frac{3 f(\kappa)}{2(\kappa^2-1)}\right) \frac{z^2}{R_z^2}-\varepsilon_{dd} f(\kappa) \bigg], \nonumber
\end{align}
where  the function $ f(\kappa)= [(1+2\kappa^2)/(1-\kappa^2)]- [(3\kappa^2 \arctan(1-\kappa^2)^{1/2})/(1-\kappa^2)^{3/2}) ] $
is monotonically decreasing, has asymptotic values $f(0)=1$ and $f (\infty) =-2$ and vanishes for $\kappa=R_{\rho}/R_z=1$
with $R_{\rho}$ and $R_z$ being the TF radii. 
The central density $n_0= 15 N/ (8\pi R_{\rho}^2 R_z)$ can be obtained via the normalization condition.
The TF Eqs.(\ref{TF:eq}) which admits also an inverted-parabola, constitute a natural extension of that obtained earlier in Refs. \cite{Dell,Dell1} within the usual GP equation. 
In the absence of the DDI, they reduce to our TDHFB equations describing nondipolar Bose polarons \cite {Boudj,Boudj1, Boudj2, Boudj3}.

When setting $\gamma=0$, the impurity and the condensate become completely decoupled hence,  $n_c=n_{c0} $ and $\tilde m=\tilde m_0$. 
For $\gamma \neq 0$, Eqs.(\ref {TF:eq1}) and (\ref {TF:eq2}) reveal that for repulsive/attractive interactions, 
the condensed and the anomalous densities are depressed/increased from their decoupled values $n_{c0} $ and $\tilde m_0$.
Therefore, the validity of our approach requires the conditions:  $n_{c0} \gg \gamma G n_I$ and $\tilde m_0 \gg \gamma (\beta-2) n_I$.
This allows to reach the weak impurity-boson  coupling regime which assumes that 
the impurity  density is much smaller than the condensed and anomalous densities.

From now on we restrict ourselves to the weakly-coupled regime. After inserting $\Psi_{dd}$ of Eq.(\ref{DDIp}) into the TF Eqs.(\ref{TF:eq1}),
we find to zeroth order
\begin{equation}\label{chmD}
\mu_B = g_B n_0 \left[ \frac{n_c}{n_0} a_1- a_2\,\varepsilon_{dd} f(\kappa) \right].
\end{equation}
This expression points out that the chemical potential associated with the condensate is decreasing  with increasing the DDI and the anomalous fluctuations.
For $\tilde m/\Phi^2 \ll 1$, one has  $a_1=a_2=1$ and thus, Eq.(\ref{chmD}) simplifies to the standard chemical potential \cite{Dell,Dell1}.

\begin{figure} 
\centerline{
\includegraphics[scale=0.45]{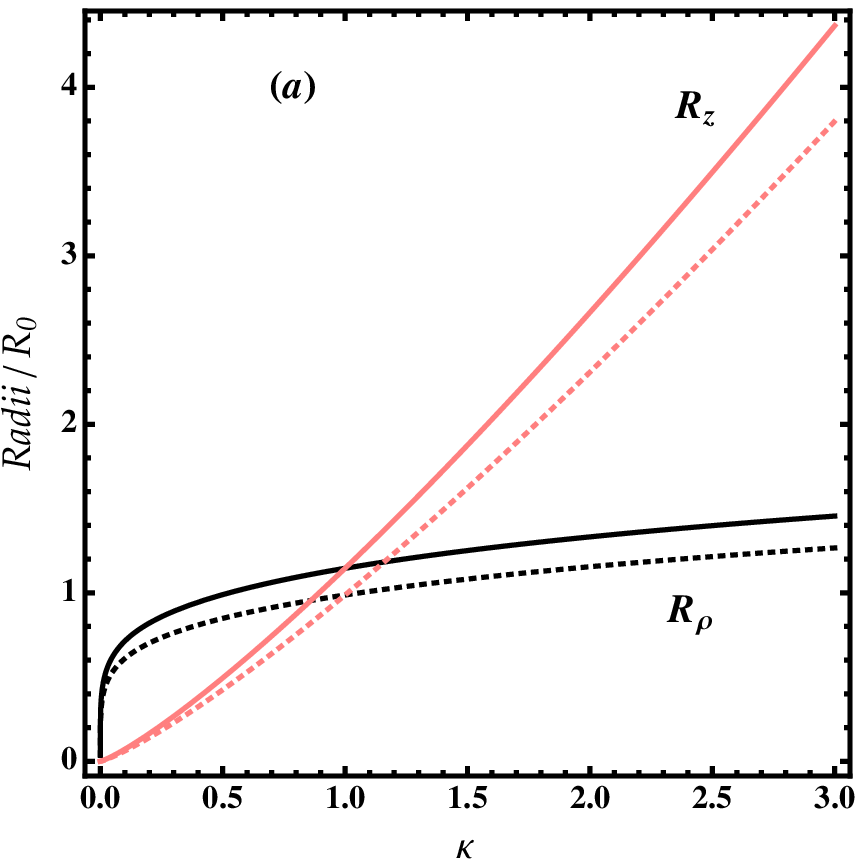}
\includegraphics[scale=0.45]{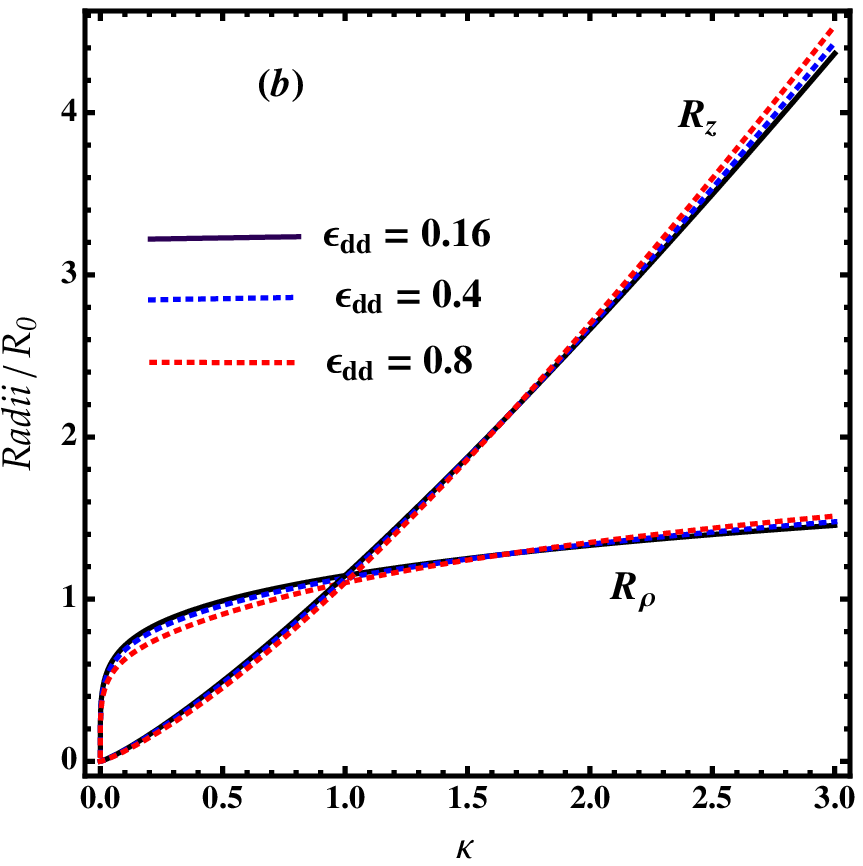}}
 \caption{ (Color online) The TF radii $R_{\rho}$ and $R_z$ of the condensate as a function of $\kappa$.
(a) Effects of the impurity for $\epsilon_{dd}=0.4$, and $N_c/N=0.9$. Soild lines: with impurity. Dotted lines: without impurity. 
(b) Effects of the DDI for $N_c/N=0.9$.
Here $R_0=[15 g_B N /4\pi m\omega_{B \rho}^2]^{1/5}$.}
\label{Rad} 
\end{figure}

The TF radii $R_{\rho}$ and $R_z$ of the condensate are obtained from (\ref{TF:eq1}) by comparing the coefficients of $\rho^2$ and $z^2$ \cite{Dell1}:
\begin{equation}\label{Radc}
R_{\rho}^c = \kappa R_z^c= \left[\frac{15 g_B N }{4\pi m\omega_{B \rho}^2} \kappa \, h_c\right]^{1/5},
\end{equation}
where $h_c=  (N_c/N) a_1+ a_2\,\varepsilon_{dd} [3 \kappa^2 f(\kappa)/2(1-\kappa^2) -1]$.
For $a_1=a_2= 1$, Eqs.(\ref{Radc}) become identical to those obtained in Ref.\cite{Dell1}.
For $\varepsilon_{dd}=0$, one recovers the usual TF radii.
Equation (\ref{Radc}) clearly shows that the radii $R_{\rho}^c$ and $ R_z^c$ are increasing with decreasing the condensed fraction $N_c/N$.
The behavior of $R_{\rho}^c$ and $ R_z^c$ with and without impurity is displayed in Fig.\ref{Rad}.a.
We see that the presence of the impurity leads to lower the TF radii. 
Figure \ref{Rad}.b. shows that for $\kappa \lesssim 1.5$, the DDI tends to reduce $R_{\rho}^c$ and $ R_z^c$. 
Whereas, for $\kappa > 1.5$, the TF radii rise with the relative interaction strength $\varepsilon_{dd}$. 

Following the same procedure, we get for the TF radii of the anomalous component 
\begin{equation}\label{Radm}
R_{\rho}^{\tilde m} = \kappa R_z^{\tilde m}= \left[\frac{15 g_B N }{4\pi m\omega_{B\rho}^2} \kappa \, h_{\tilde m} \right]^{1/5},
\end{equation}
where $h_{\tilde m}=  (\tilde M/N) a_3+ a_4\,\varepsilon_{dd} [3 \kappa^2 f(\kappa)/2(1-\kappa^2) -1]$, and $\tilde M= \int \tilde m \, d \mathbf r$. 
The same behavior persists for $R_{\rho}^{\tilde m}$ in terms of  $\varepsilon_{dd}$,  $\tilde M/N$, and $\beta$.

\section{ Dynamics and breathing oscillations of a trapped impurity} \label{SLS}

In this section we analyze effects of the impurity-BEC  interaction strength, the DDI and the anomalous correlations on the
breathing modes of the widths of a harmonically trapped impurity interacting with a trapped BEC using variational method and numerical scheme based on our TDHFB equations.


Substituting Eqs.(\ref{TF:eq}) into Eq.(\ref{E:eq4c}), one obtains the extended self-focusing nonlinear Schr\"{o}dinger equation (NLSE)
 \begin{equation}\label{E:eq7}
i\hbar \dot{\Phi}_I= \left[ h_I^{sp} +g_B\gamma n_0 \left(1-\frac{\rho^2}{R_{\rho}^2}-\frac{z^2}{R_{z}^2}\right) - g_B \gamma^2 \lambda n_I \right] \Phi_I,
\end{equation}
where $\lambda=\beta -2 +G $, $1/R_{\rho}^2= (R_{\rho}^{c2}+R_{\rho}^{\tilde m 2})/R_{\rho}^{c2}R_{\rho}^{\tilde m 2} $, 
and $1/R_z^2= (R_z^{c2}+R_z^{\tilde m 2})/R_z^{c2}R_z^{\tilde m 2} $.
The last term in Eq.(\ref{E:eq7}) describes the back action of the condensate on the impurity which is caused by the response of the condensate density on
the presence of the impurity due to their interaction.
This equation is appealing since it enables us to study, in useful manner, the effects of normal and anomalous correlations, 
the DDI, and the impurity-host interaction on the dynamical properties of the impurity state. 
For $\tilde m=\tilde n=0$, one recovers the standard NLSE. 

To estimate the widths of the harmonic trapping of the impurity in the presence of the envirement fluctuations,
we use Gaussian variational ansatz for the impurity wavefunction, 
with an axial size $\sigma_z$ and a radial size $\sigma_{\rho}$ that we take as variational parameters:
\begin{align} \label{IWF}
\Phi_I = \sqrt{\frac{1}{\pi^{3/2}\sigma_\rho \sigma_z }} \exp\left[-\frac{\rho^2}{2 \sigma_{\rho}^2}-\frac{z^2}{2 \sigma_z^2}- i \Theta_\rho \rho^2 - i\Theta_z z^2\right],
\end{align}
where $\Theta_\rho $ and $\Theta_z $ stand for the radial and axial phases.

The Lagrangian density corresponding to Eq.(\ref{E:eq7}) reads:
\begin{align} \label{E:eq8}
\mathcal{L}&=  \frac{i\hbar}{2}(\Phi_I^* \dot{\Phi}_I-\Phi_I \dot{\Phi}_I^*)-{\frac{\displaystyle\hbar^2}{\displaystyle 2m_I}}\Phi_I^* \Delta \Phi_I \\
&-\left[ U_I +g_B \gamma  n_0 \left(1-\frac{\rho^2}{R_{\rho}^2}-\frac{z^2}{R_{z}^2}\right)-g_B \gamma^2 \lambda|\Phi_I|^2\right] |\Phi_I|^2. \nonumber
\end{align}
\begin{figure}
\includegraphics[scale=0.45] {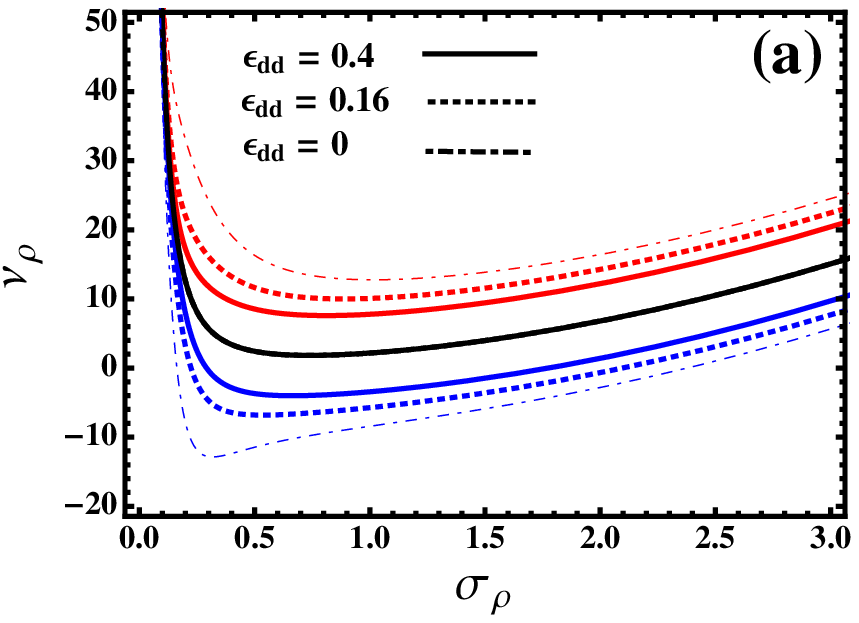}
\includegraphics[scale=0.45] {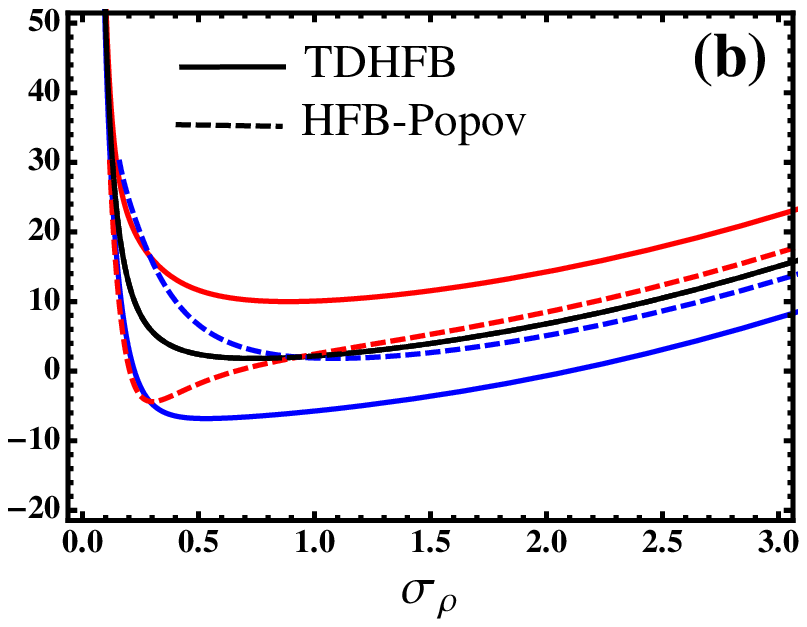}
\includegraphics[scale=0.45] {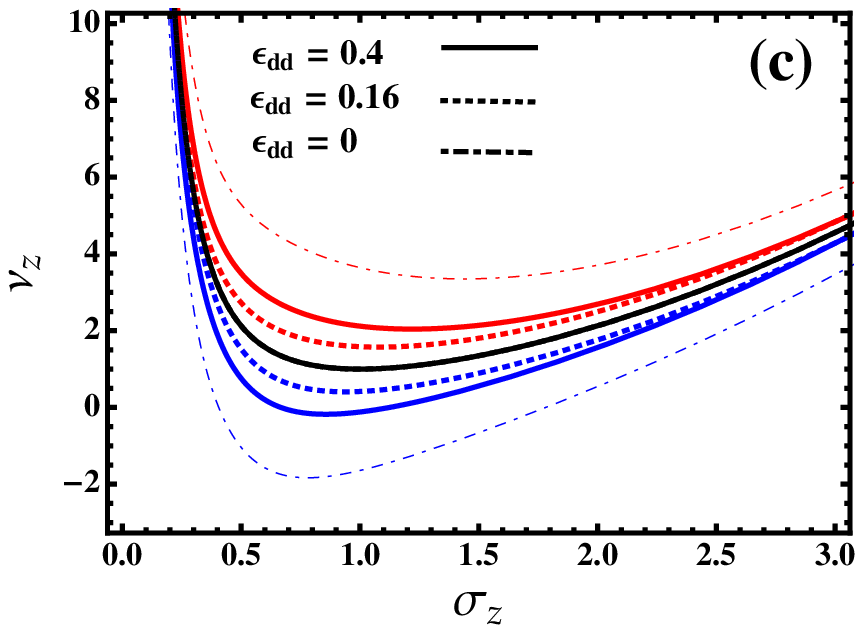}
\includegraphics[scale=0.45] {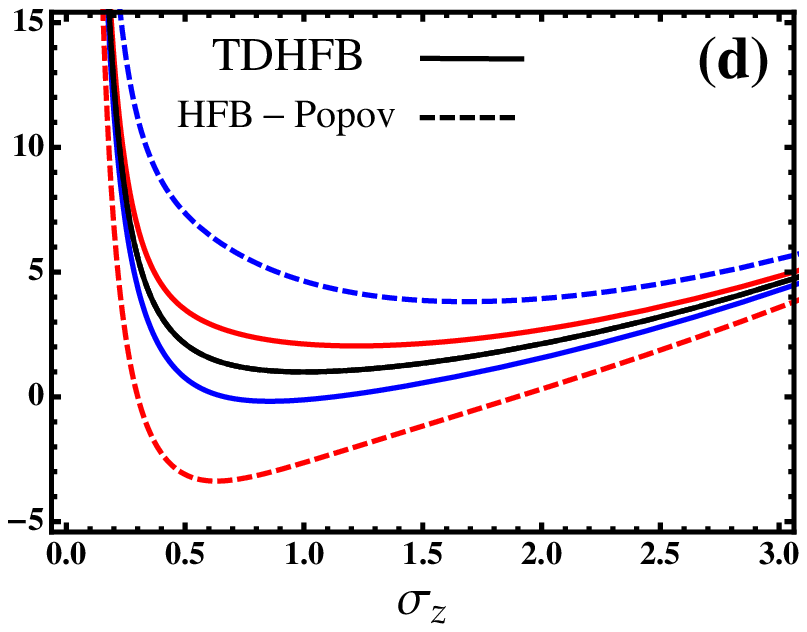}
\caption{(Color online) The effective radial (top panel) and axial (bottom panel) potentials from Eqs.(\ref{E:eq11}) and (\ref{E:eq12}) for several values of $\gamma$. 
Red lines: $\gamma=-2$. Black lines: $\gamma=0$. Blue lines: $\gamma=2$. 
(a) and (c) Effects of DDI. (b) and (d)  Comparison between HFB-Popov approximation ($\tilde m=0$) and the TDHFB predictions for $\varepsilon_{dd}=0.4$. 
Parameters are: $N=10^3$, $a_B= 100 \,a_0$ with $a_0$ being the Bohr radius, $\omega_{B \rho}=2\pi \times 500$ Hz, $\lambda_B=\lambda_I=5$,  
$\alpha=2$, and $\Omega_{\rho}=0.2$.}
 \label {EfPot}
\end{figure}
Variational equations of motion can be derived from the Euler-Lagrange equations $\frac{\partial}{\partial t}\left(\frac{\partial L}{\partial\dot{q}}\right)-\frac{\partial L}{\partial q} $,
where $L=\int \mathcal{L} d {\bf r}$ and $q$ stands for the variational parameters we introduced above.  The radial and axial phases can be written as 
$\Theta_\rho= \frac{-m_I}{2 \hbar}\frac{\dot{\sigma}_\rho}{\sigma_\rho}$, and  $\Theta_z=\frac{-m_I}{2 \hbar}\frac{\dot{\sigma}_z}{\sigma_z}$.
After having eliminating the phase from ordinary lagrangian, 
we obtain useful equations for the radial and axial widths: 
\begin{align} 
\ddot{\tilde{\sigma}}_\rho =\frac{2}{N_I\hbar \omega_{Iz}} \frac{\partial}{\partial \tilde{\sigma}_\rho } {\cal V}_{\rho}, \label{E:eq13} \\
\ddot{\tilde{\sigma}}_z =\frac{2}{N_I\hbar \omega_{Iz}} \frac{\partial}{\partial \tilde{\sigma}_z } {\cal V}_z, \label{E:eq14}
\end{align}
where the radial $ {\cal V}_{\rho}$ and axial ${\cal V}_z$ effective potentials are given, respectively as 
\begin{align}\label{E:eq11}
 {\cal V}_{\rho}=\frac{1}{\tilde{\sigma}_\rho^2}+ \frac{\tilde{\sigma}_\rho^2}{\lambda_I^2} - \frac{g_B \gamma^2 \lambda I_0}{\tilde{\sigma}_\rho^2 \tilde{\sigma}_z}
+g_B \gamma I_1 {\cal J}_1 (\tilde{\sigma}_\rho, \tilde{\sigma}_z), 
\end{align}
and 
\begin{equation}\label{E:eq12}
{\cal V}_z = \frac{1}{2 \tilde{\sigma}_z^2}+\frac{\tilde{\sigma}_z^2}{2}- \frac{g_B \gamma^2 \lambda I_0}{\tilde{\sigma}_\rho^2 \tilde{\sigma}_z} +\frac{g_B\gamma I_1}{2}  
{\cal J}_2  (\tilde{\sigma}_\rho, \tilde{\sigma}_z),
\end{equation}
where  
\begin{align} 
{\cal J}_1 (\tilde{\sigma}_\rho, \tilde{\sigma}_z)=\frac{\erf{(\tilde{\sigma}_z)}}{\tilde{R}} \left[(e^{\tilde{\sigma}_\rho}+1)\frac{\tilde{\sigma}_\rho^2}{\kappa^2}+\frac{e^{\tilde{\sigma}_\rho}\tilde{\sigma}_z^2}{2}\right]
-\frac{\tilde{\sigma}_z}{\sqrt{\pi\tilde{ R}}} e^{\tilde{\sigma}_z} e^{\tilde{\sigma}_\rho}, \nonumber
\end{align}
\begin{align} 
{\cal J}_2  (\tilde{\sigma}_\rho, \tilde{\sigma}_z) &= \erf{(\tilde{\sigma}_z)}+\frac{2}{\tilde{R}} \erf{(\tilde{\sigma}_z)} (e^{\tilde{\sigma}_\rho}-1)
\left(\frac{\tilde{\sigma}_\rho^2}{\kappa^2}+\frac{\tilde{\sigma}_z^2}{2}\right) \nonumber\\
&+ \frac{2}{\sqrt{\pi\tilde{ R}}} \tilde{\sigma}_z  e^{\tilde{\sigma}_z}(1-e^{\tilde{\sigma}_\rho}), \nonumber
\end{align}
$I_0=1 /(\hbar \omega_{Iz} \sqrt{2}\pi^{3/2} l_{Iz}^3)$, and $ I_1=2n_0/\hbar \omega_{Iz}$.
Here we have introduced the  dimensionless quantities
$\tilde{\sigma}_\rho =\sigma_\rho/l_I, \, \tilde{\sigma}_z= \sigma_z/l_I, \, \tilde{t}=\omega_{Iz} t, \, \tilde{R}=2 R_z^2/l_I$
with $l_I=\sqrt{\hbar/m_I \omega_{Iz}}$ being the impurity's harmonic-oscillator length.
The first two terms in Eqs.(\ref{E:eq11}) and (\ref{E:eq12}) describe the quantum pressure and the harmonic trapping of the impurity. 
The third contribution is second-order in $\gamma$ and represents self-trapping effects. 
It originates from the deformation of the condensate due to the interaction with the impurity. 
The above equations (\ref{E:eq11}) and (\ref{E:eq12})  show that for  $\gamma=0$ i.e BEC without impurity, 
the effective potentials are much independent on the DDI and on the anomalous effects. This is clearly visible in Fig.\ref{EfPot}.

The radial $ {\cal V}_{\rho}$ and axial ${\cal V}_z$ effective potentials have a local minimum indicating the localization of the impurity
in the condensate as is seen in Fig.\ref{EfPot}. 
The depth and the width of such a local minimum are strongly affected by the DDI, impurity-BEC interactions and the anomalous fluctuations.
For instance, for repulsive coupling, the impurity becomes less localized in the ground state for increasing DDI while the situation is inverted
for impurity-BEC interactions (see Figs.\ref{EfPot}.a and c). 
An interesting remark is that in the radial direction, the potential $ {\cal V}_{\rho}$ is flattened near its minimum for both repulsive and attractive impurity-boson coupling 
(see Figs.\ref{EfPot}.b and d) due to the presence of the anomalous fluctuations. 
In the axial direction, the location and height of the local minima are modified to some extent 
because of the external trapping force \cite{THJ}.

\begin{figure}
\includegraphics[scale=0.66] {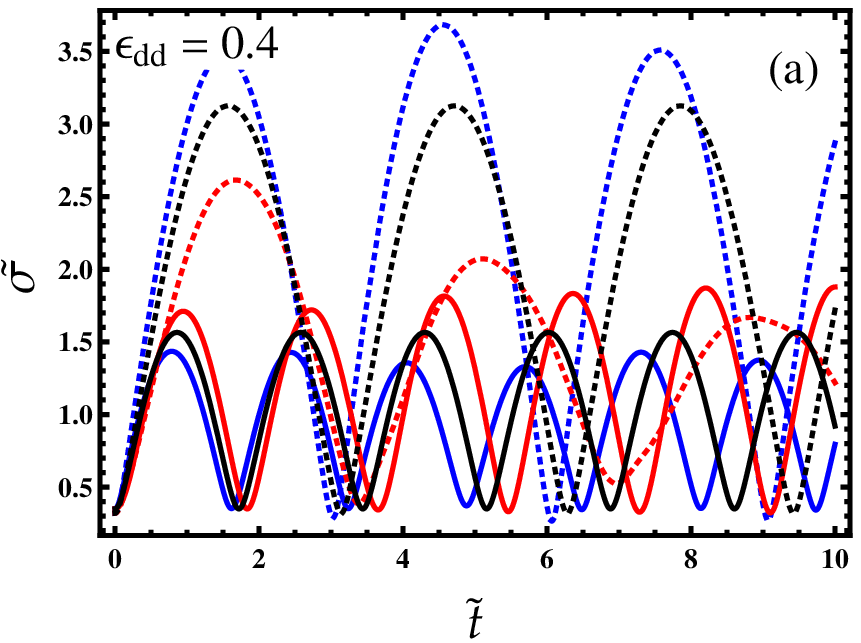}
\includegraphics[scale=0.65] {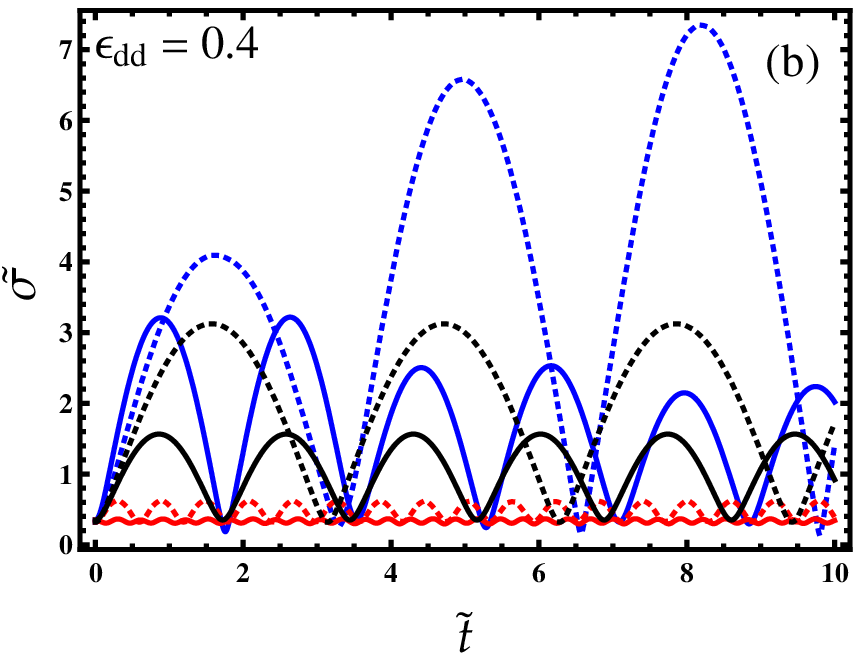}
\caption{(Color online) Radial ( $\tilde \sigma_\rho$: dotted lines) and axial ($\tilde \sigma_z$: solid lines) widths as a function of time $\tilde t$.
Here (a) : TDHFB results. (b) HFB-Popov calculations. 
Red lines: $\gamma=-2$. Black lines: $\gamma=0$. Blue lines: $\gamma=2$.
Parameters are the same as in Fig.\ref{EfPot}.}
\label {widt}
\end{figure}

Equations of motion for the radial and axial widths  (\ref{E:eq13}) and (\ref{E:eq14}) of the impurity are differential equations, which can be solved by using an appropriate numerical scheme.
The solutions are presented in  Fig.\ref {widt}.
We find that in the presence of the anomalous fluctuations, the oscillation amplitude and width of $\tilde \sigma_{\rho}$ and of $\tilde \sigma_z$ are
increasing/decreasing with increasing/decreasing the interspecies interactions $\gamma$  as is seen in Fig.\ref {widt}.a
which is in agreement with experiments of Ref.\cite{Cat1} achieved for nondipolar polarons.  
For $\tilde m =0$ and $\gamma >0$, the impurity oscillates slower and with a large amplitude 
($\sim $ 2 times bigger than that predicted by the TDFHB theory) as is depicted in Fig.\ref {widt}.b.
Fast periodic oscillations with small amplitudes are observed  when $\gamma < 0$.
 Strong oscillation strength means that the impurity oscillations are confined inside the bath. 
Remarkably, in the axial direction, the impurity gains kinetic energy and vibrate faster and stronger compared to the radial direction 
irrespective of the presence of the impurity and the anomalous correlations. 
This is most probably due to the anisotropy of the DDI.


\begin{figure}
\includegraphics[scale=0.42] {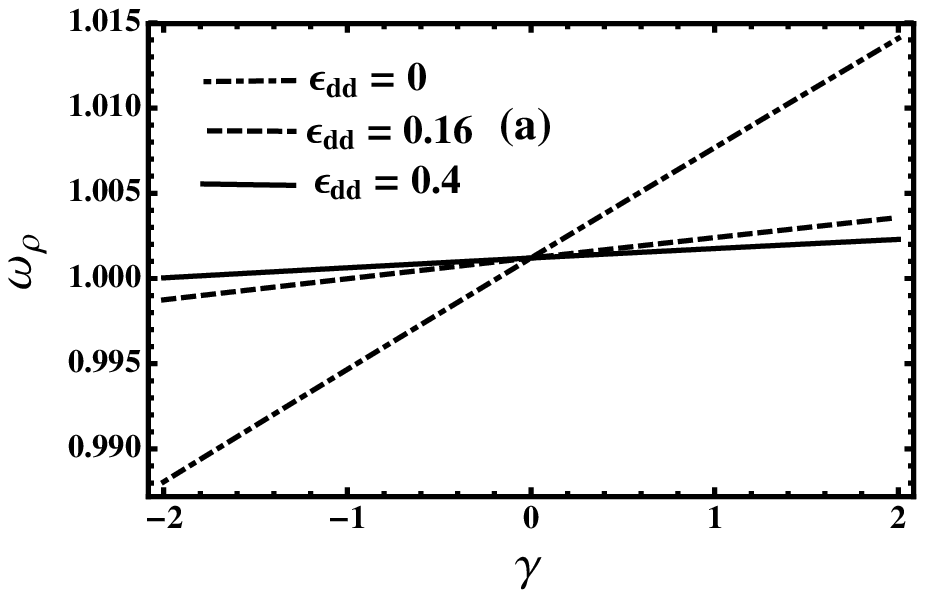}
\includegraphics[scale=0.42] {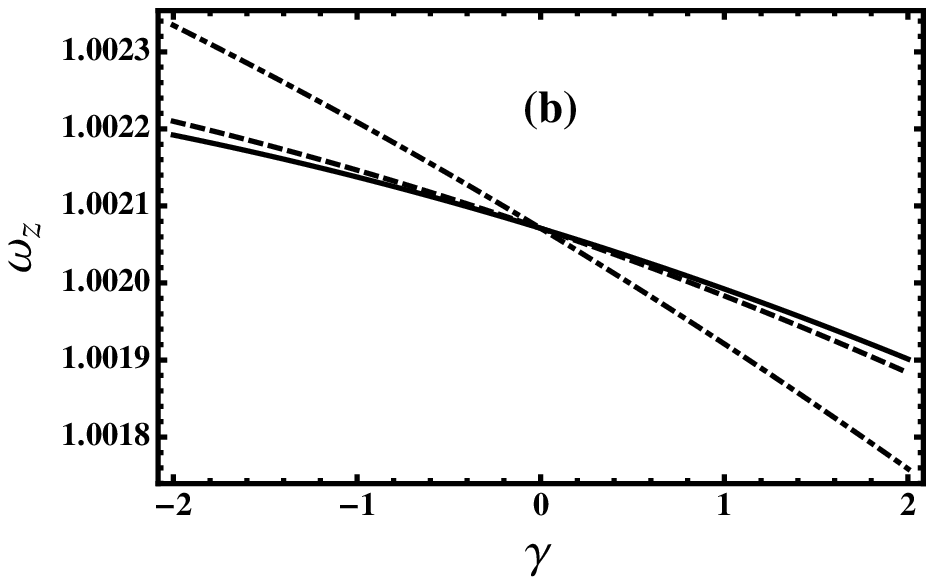}
\includegraphics[scale=0.42] {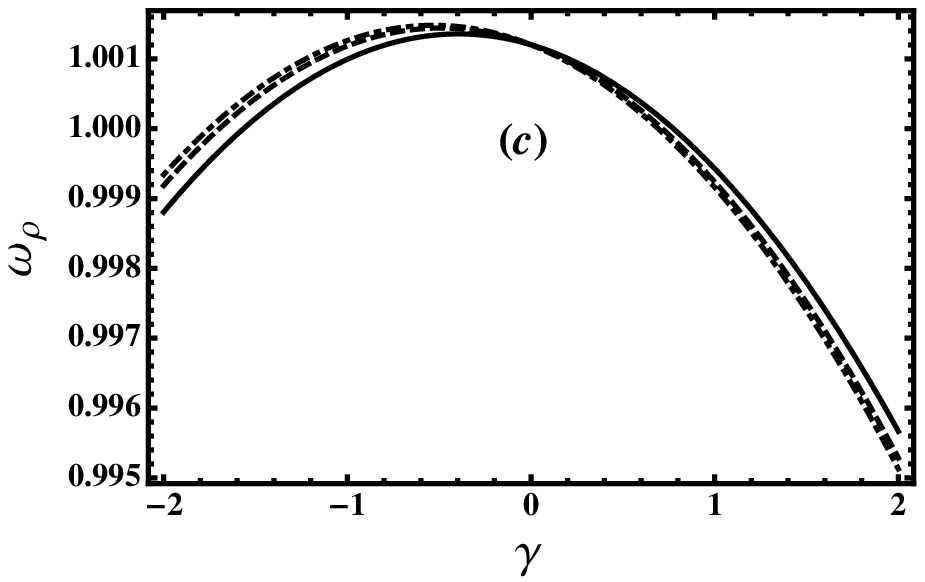}
\includegraphics[scale=0.42] {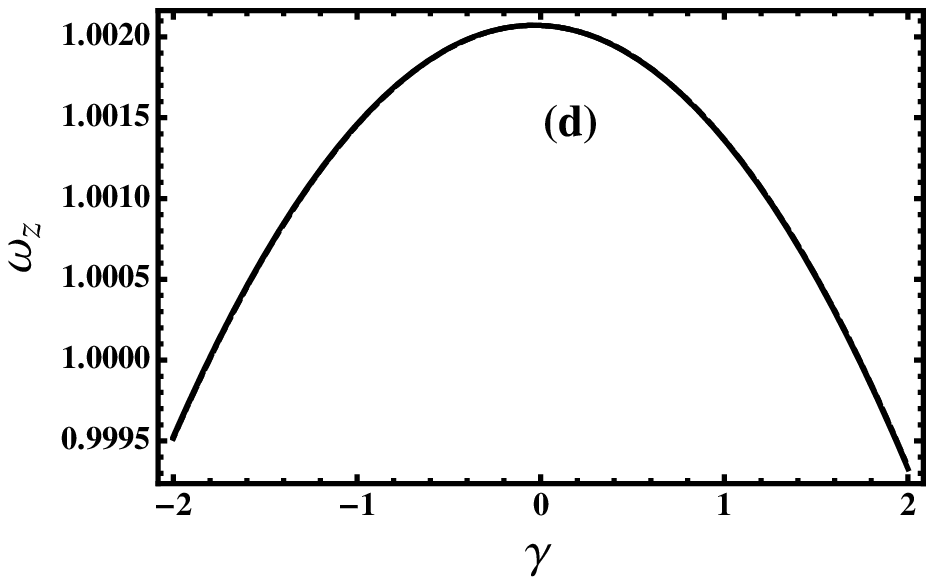}
 \caption{The radial and axial frequencies of the breathing modes as a function of $\gamma$.
(a) and (b) for different values of $\varepsilon_{dd}$. (c) and (d) HFB-Popov predictions. 
Parameters are the same as in Fig.\ref {EfPot}.}
\label {colmod}
\end{figure}

The breathing modes frequencies of the width can be obtained by linearizing Eq.(\ref{E:eq13}) and expanding the radial and axial widths of the impurity as
 $\tilde{\sigma}_\rho(t)=\tilde{\sigma}_{\rho0}+\delta\tilde{\sigma}_\rho(t)$, and $\tilde{\sigma}_z(t)=\tilde{\sigma}_{z0}+\delta\tilde{\sigma}_z(t)$, 
where $\tilde{\sigma}_{\rho 0}$ and $\tilde{\sigma}_{z0}$ are the equilibrium radial and axial widths, respectively, and 
$\delta\tilde{\sigma}_\rho(t)/\tilde{\sigma}_{\rho0} \ll 1$, and $\delta\tilde{\sigma}_z(t)/\tilde{\sigma}_{z0} \ll 1 $. 
Neglecting higher order terms, we obtain
\begin{equation}\label{FR1}
w_\rho= \Bigg( \frac{2}{\lambda_I^2}+\frac{6}{\sigma_{\rho0}^4}-\frac{6 g_B \gamma^2 \lambda I_0}{\sigma_{\rho0}^4 \sigma_{z0}}
+g_B \gamma I_1 \frac{\partial^2 {\cal J}_2  (\tilde{\sigma}_{\rho0}, \tilde{\sigma}_{z0})}{\partial \sigma_{\rho0}^2} \Bigg)^{1/2},
\end{equation}
and
\begin{equation}\label{FR2}
w_z =\Bigg( 1+\frac{3}{\sigma_{z0}^4}-\frac{2 g_B \gamma^2 \lambda I_0}{\sigma_{\rho0}^2 \sigma_{z0}^3}+
\frac{g_B \gamma  I_1}{2}  \frac{\partial^2 \mathcal{J}_1 (\tilde{\sigma}_{\rho0}, \tilde{\sigma}_{z0})}{\partial \sigma_{z0}^2} \Bigg)^{1/2}.
\end{equation}
Equations (\ref{FR1}) and (\ref{FR2}) are appealing since they describe the impurity's breathing modes of the widths in terms of the DDI, impurity-boson interaction, 
the anomalous correlation and the trapping frequency at the same time.

In Fig.\ref{colmod},  we plot our predictions for radial and axial breathing modes as a function of $\gamma$ and compare the results
with the HFB-Popov findings.  
In the radial direction, the frequencies of the breathing modes $w_\rho$ slightly increase with $\varepsilon_{dd}$ for $\gamma <0$ while they 
decrease with $\varepsilon_{dd}$ for $\gamma > 0$ (see Fig.\ref{colmod}.a). 
For fixed value of DDI (say $\varepsilon_{dd}=0.4$), the frequencies $w_\rho$ augment with  $\gamma $ in the whole spectrum of impurity-BEC interaction
regimes from attractive to repulsive.
In the axial direction however, the opposite behavior is observed as a result of the anisotroy of the DDI (see Fig.\ref{colmod}.b).
Moreover, it is interesting to observe that the TDHFB predictions and the HFB-Popov results differ from each other quite considerably
owing to the important contribution of the anomalous density as is shown in Figs.\ref{colmod}.c and d. 
We see in particular that such frequencies in both radial and axial directions rise with $\gamma$ until they reach their maximal value at $\gamma=0$,
then they start to decrease for repulsive impurity-BEC interaction. 
We remark also that in the absence of the anomalous density, the impact of the DDI on the breathing modes is not important.

\section{Conclusion} \label {Conc}

In this paper we investigated in details the effects of the normal and anomalous quantum fluctuations on the dynamics of dipolar Bose polarons
within the framework of the TDHFB theory. 
We showed that the impurity distorts the condensate and the anomalous component for both repulsive and attractive impurity-host interactions. 
For repulsive interations, the depth of the dip structure caused by the impurity becomes more profound for large DDI. 
The time evolution of the condensate and the anomalous component in the presence of the impurity has been deeply analyzed. 
We found that the external trapping potential and the interspecies interactions may significantly 
modify the oscillations of the width and the center-of-mass of such quantities.
Moreover, corrections due to quantum fluctuations to the radii and chemical potentials of both the condensate and the 
anomalous component have been calculated in the weak coupling limit using the TF approximation.

In addition, we studied the properties of a localized impurity for both repulsive and attractive contact impurity-boson interactions and dipolar boson-boson interaction. 
Analytical results for the width and breathing frequencies of the impurity stemming from our generalized NLSE have been obtained using a variational approach.
We pointed out that the interplay of the DDI and the impurity-BEC interactions play a key role in 
the dynamics and in the breathing oscillations frequencies of the impurity. 
We compared our findings for the dynamics and the breathing mode frequency of the impurity with HFB-Popov predictions. 
Results show that, to some extent, both theories conflict with each other due to the effects of the anomalous fluctuations.


Our analysis can be extended to the finite temperature using the complete TDHFB theory where Eq.(\ref{Inv1}) could be rewritten  as 
$ \left[2\tilde n({\bf p, r})+1\right]^2- 4|\tilde m({\bf p, r})|^2= \text {coth}^2\left(\varepsilon({\bf p, r})/2T \right)$ \cite{Boudj11}.
In such a case, one can expect that the impurity becomes less localized. 
The breathing oscillations frequencies could be also shifted in particular at intermediate temperatures where the anomalous density  reaches its maximal value \cite{Boudjbook}. 
Our approach still valid for studying the dynamics of dipolar polarons with tilted dipoles.
The findings of the present work provide a good starting point for exploring effects of the impurity in dipolar quantum droplets.

\section{Acknowledgments}
We gratefully acknowledge Stefano Giogini and Georg Brunn for comments on this paper.





\end{document}